\documentclass[aps,pra,reprint,floatfix,superscriptaddress]{revtex4-1}

\usepackage{amsmath,amssymb}
\usepackage{graphicx}
\usepackage{bm}
\usepackage{cases}

\usepackage[breaklinks]{hyperref} 

\usepackage{framed,color}
\definecolor{shadecolor}{rgb}{1,0.8,0.3}

\begin{document}

\title{Effective self-similar expansion for the Gross-Pitaevskii equation} 
\author{Michele Modugno}
\affiliation{\mbox{Depto. de F\'isica Te\'orica e Hist. de la Ciencia, Universidad del Pais Vasco UPV/EHU, 48080 Bilbao, Spain}}
\affiliation{IKERBASQUE, Basque Foundation for Science, Maria Diaz de Haro 3, 48013 Bilbao, Spain}
\author{Gianni Pagnini}
\affiliation{IKERBASQUE, Basque Foundation for Science, Maria Diaz de Haro 3, 48013 Bilbao, Spain}
\affiliation{BCAM - Basque Center for Applied Mathematics, Alameda de Mazarredo 14, 48009 Bilbao, Spain}
\author{Manuel Angel Valle-Basagoiti}
\affiliation{\mbox{Depto. de F\'isica Te\'orica e Hist. de la Ciencia, Universidad del Pais Vasco UPV/EHU, 48080 Bilbao, Spain}}


\begin{abstract}
We consider an effective scaling approach for the free expansion of a one-dimensional quantum wave packet, 
consisting in a \textit{self-similar} evolution to be satisfied \textit{on average}, i.e. by integrating over the coordinates. A direct comparison with the  solution of the Gross-Pitaevskii equation shows that the effective scaling reproduces with great accuracy the exact evolution -- the actual wave function is reproduced with a fidelity close to one -- for arbitrary values of the interactions. This result represents a proof-of-concept of the effectiveness of the scaling ansatz, which has been used in different forms in the literature but never compared against the exact evolution.
\end{abstract}

\maketitle

\section{Introduction}

The scaling approach is a powerful method for reducing the complexity of partial differential equations in cases where the evolution of the system is characterized by a self-similar behavior which is captured by a simple rescaling of the coordinates. In this case, the dynamical equations for a $d$-dimensional system are replaced by $d$ ordinary differential equations (with respect to time) for the scaling parameters. This approach has been successfully employed for describing  the collective excitations and the free expansion of Bose-Einstein condensates of interacting atomic gases in different geometries (for which exact solutions exists both in the noninteracting limit and in the Thomas-Fermi regime, where interactions dominate over the kinetic energy)\cite{castin1996,kagan1996,dalfovo1999,timmermans2000,brazhnyi2003,kamchatnov2004,egusquiza2011}, the expansion of a one-dimensional Bose gas (in the deep Thomas-Fermi regime and in the Tonks-Girardeau regime of impenetrable bosons) \cite{pedri2003}, of a superfluid Fermi gas \cite{giorgini2008,schafer2010,egusquiza2011}, and of a thermal cloud \cite{bruun2000}.

Remarkably, an \textit{effective scaling approach} has also been used by different authors as an approximate solution for the evolution of both bosonic and fermionic density distribution, including the collective excitations of a trapped Bose gas \cite{guery-odelin2002}, the expansion of an interacting Fermi gas \cite{menotti2002}, and the expansion of a quantum degenerate Bose-Fermi mixture \cite{hu2003}. This effective scaling consists in using a self-similar ansatz for the evolution of the system in the hydrodynamic regime, and -- when the scaling is not an exact solution -- imposing the corresponding hydrodynamic equations to be satisfied \textit{on average}, by integrating over the coordinates. However, this effective approach has never been tested against the exact solution of the corresponding equations. As a proof-of-concept, here we consider a simple problem that admits a direct comparison between the effective and the exact evolutions of the system.  

In particular, we shall consider the free expansion of a quasi-one-dimensional Bose-Einstein condensate that is initially prepared in the ground state of a harmonic trap (at zero temperature). Here quasi-one-dimensional refers to an elongated condensate the transverse degrees of freedom of which are frozen during the entire evolution due to a large radial confinement (see e.g. \cite{pedri2003}). It is worth remarking that though for a homogeneous system in one dimension one cannot have Bose-Einstein condensation in the thermodynamic limit \cite{dalfovo1999}, in the presence of harmonic confinement the system can exhibit a macroscopic occupation of the lowest-energy state and the state of the system can be indeed described by a quasi-condensate (a condensate with fluctuating phase) or a true condensate \cite{petrov2000}. We shall consider the latter situation, by solving the corresponding nonlinear Gross-Pitaevskii equation \cite{dalfovo1999}, that is valid in the weakly interacting, mean-field regime. By applying the effective scaling approach mentioned above, we obtain an equation for the scaling parameter interpolating between the noninteracting and the Thomas-Fermi limits (where the self-similarity is exact). We show that this approach is indeed very accurate in reproducing the exact evolution, for arbitrary values of the interactions. It is worth noticing that these mean-field results contrast with the case of a strongly interacting one-dimensional Bose gas, for which the self-similarity is explicitly violated in the crossover between the mean-field Thomas-Fermi regime the Tonks-Girardeau regime \cite{ohberg2002,pedri2003}.

The paper is organized as follows. In Sect. \ref{sec:model} we discuss the effective scaling ansatz in the hydrodynamic formulation of the Gross-Pitaevskii equation, and derive the ordinary differential equation for the scaling parameter. Then, in Sect. \ref{sec:results} we discuss the numerical solution and the asymptotic limit of the scaling equation, for different values of the interaction strength,
ranging from the noninteracting limit to the Thomas-Fermi limit. Here we also analyze the fidelity of this effective approach in reproducing the exact quantum evolution of the system, as dictated by the Gross-Pitaevskii equation, finding that this method is indeed very accurate for arbitrary values of the interactions. Final considerations are drawn in the conclusions.

\section{Model}
\label{sec:model}

In this paper we shall consider a quasi-one-dimensional Bose-Einstein condensate -- described by the wave function $\psi(x,t)$ -- that expands in free space according to the Gross-Pitaevskii equation
\begin{equation}
i\hbar\partial_{t}\psi=\left[ -\frac{\hbar^2}{2m}\partial_{x}^{2} + g|\psi|^{2} \right]\psi,
\label{eq:gpe}
\end{equation}
where $m$ is the particle mass, and the coefficient $g$ of the nonlinear term, representing the strength of the inter-particle interaction, is here assumed to be non negative (repulsive interaction).
The above equation is equivalent to the following set of hydrodynamic equations \cite{dalfovo1999}
\begin{numcases}{}
\label{eq:continuity}
{\partial_{t} n} + \partial_{x} \left(n{v}\right) = 0  \\
m{\partial_{t}v} + \partial_{x}  \left[ P(x,t) +\frac{1}{2} m{v}^2 + gn \right] = 0,
\label{eq:euler}
\end{numcases}
which is obtained by posing $\psi=\sqrt{n}e^{i S}$, $n=|\psi|^{2}$, and $v=(\hbar/m)\partial_{x} S$ \cite{stringari1996} (the Madelung formulation \cite{madelung1926,madelung1927}), where we have defined
\begin{equation}
P(x,t)=-\frac{\hbar^2}{2m}\frac{1}{\sqrt{n}}\partial_{x} ^2\sqrt{n}.
\end{equation}
Here $n(x,t)\equiv|\psi(x,t)|^{2}$ represents the particle density, and $v(x,t)$ represents the corresponding velocity field.
In the following, we shall refer to the case in which the condensate is initially prepared in the ground state of a harmonic potential of frequency $\omega_{0}$, $V(x)=(1/2)m\omega_{0}^{2}x^{2}$, as in typical experimental setups \cite{dalfovo1999}. In the noninteracting limit ($g=0$) and in the Thomas-Fermi regime ($g\gg1$), Eqs. (\ref{eq:continuity}) and (\ref{eq:euler}) admit exact scaling solutions of the form \cite{dalfovo1999}  
\begin{align} 
\label{eq:scaling-n}
n(x,t)&=\frac{1}{\lambda(t)}n_0\left(\frac{x}{\lambda(t)}\right),
\\
v(x,t) &= \frac{\dot{\lambda}(t)}{\lambda(t)}x,
\label{eq:scaling-v}
\end{align}
where $n_{0}(x)$ is the initial density distribution, and the parameter
$\lambda(t)$ satisfies $\ddot{\lambda}={\omega_{0}^{2}}/{\lambda^{n}}$, with $n=3$ or $2$ in the two limits, respectively. 
The above approach can be extended to intermediate regimes (i.e. arbitrary values of $g$), by means of an effective scaling, as considered by different authors \cite{guery-odelin2002,menotti2002,hu2003}. The main idea is to consider an ansatz in which the left member of Eq. (\ref{eq:euler}) is not vanishing at each point of space, but only after integration over the coordinates. Then, by inserting Eq. (\ref{eq:scaling-v}) in Eq. (\ref{eq:euler}), and performing a spatial integration, one gets
\begin{equation}
-\frac12 m\frac{\ddot{\lambda}}{\lambda}x^{2}=P + gn - f 
\end{equation}
with $f(t) =P(0,t) + gn(0,t)$. By introducing the rescaled coordinate $\xi=x/\lambda$, 
multiplying the above equation by $n_{0}(\xi)$, and integrating over $d\xi$, we have
\begin{equation}
f -\frac12 m\ddot{\lambda}\lambda \sigma_{0}^{2}=
\frac{E_{k}^{0}}{\lambda^{2}}
 + \frac{2E_{int}^{0}}{\lambda},
\end{equation}
where 
\begin{align}
\sigma_{0}^{2}&=\int \xi^{2}n_{0}(\xi)d\xi,
\\
E_{int}^{0}&=\frac{g}{2}\int n_{0}^{2}(\xi)d\xi ,
\\
E_{k}^{0}&=\frac{\hbar^2}{2m}\int (\partial_{\xi}\sqrt{n_{0}(\xi)})^{2}d\xi.
\end{align}
We also have
\begin{align}
P(\xi,t)
&=\frac{1}{\lambda^{2}}\frac{\hbar^2}{2m}
\left(\frac{1}{4n_{0}^{2}}(\partial_{\xi} n_{0})^{2}-
\frac{1}{2n_{0}}\partial_{\xi}^{2} n_{0}\right)
\end{align}
and 
\begin{equation}
P(0,t)=-\frac{1}{\lambda^{2}}\frac{\hbar^2}{2m}\left.
\frac{1}{2n_{0}}\partial_{\xi}^{2} n_{0}\right|_{\xi=0}
\equiv \frac{D_{0}}{\lambda^{2}}
\end{equation}

By combining all these results, one eventually arrives at the following equation for the scaling parameter
\begin{equation}
\ddot{\lambda}= \frac{A\omega_{0}^{2}}{\lambda^{3}}  + \frac{B\omega_{0}^{2}}{\lambda^{2}}
\label{eq:scaling}
\end{equation}
where $A$ and $B$ are defined as
\begin{align}
\label{eq:A}
 A&=\frac{2}{\sigma_{0}^{2}}
\frac{D_{0} -E_{k}^{0}}{m\omega_{0}^{2}},
\\
B&=\frac{2}{\sigma_{0}^{2}}
\frac{g n_{0}(0) - 2E_{int}^{0}}{m\omega_{0}^{2}}.
\label{eq:B}
\end{align}
Remarkably, they depend on the initial conditions only. It is straightforward to check that the known results for the noninteracting and Thomas-Fermi limits are correctly reproduced. Namely, for $g=0$ one has $A=1$ and $B=0$, whereas in the Thomas-Fermi limit the term $A$ is negligible and $B=1$ \cite{kamchatnov2004}.
For intermediate regimes, no explicit solutions exist, and one has to compute numerically the parameters $A$ and $B$ from the ground-state solution $\psi_{0}$ of the stationary Gross-Pitaevskii equation 
\begin{equation}
\left[ -\frac{\hbar^2}{2m}\partial_{\xi}^{2} + \frac{1}{2}m\omega_{0}^{2}\xi^{2} + g|\psi_{0}|^{2} \right]\psi_{0} = \mu\psi_{0},
\label{eq:gpes}
\end{equation}
where $\mu$ is the chemical potential
(notice that at $t=0$ there is no difference between the coordinate $x$ and $\xi$, owing to the fact that $\lambda(0)=1$).
In addition, it can be easily proved 
that the two parameters satisfy $A+B=1$, so that only one of the two is actually needed.
In fact, by posing $\psi_{0}=\sqrt{n}_{0}$ and left-multiplying by $\sqrt{n}_{0}$ Eq. (\ref{eq:gpes}) it follows that $\mu=D_{0}+g n_{0}(0)$. Inserting this expression of the chemical potential again in the previous equation, integrating over the coordinate $\xi$, and comparing the result with the expressions (\ref{eq:A}) and (\ref{eq:B}), it is straightforward to verify that $A+B=1$.

\section{Results and discussion}
\label{sec:results}

\begin{figure}[tbp]
\centerline{\includegraphics[width=0.98\columnwidth]{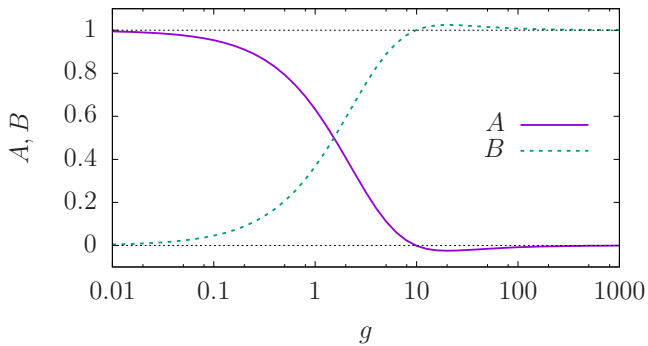}}
\caption{Behavior of the parameter $A$ and $B$ defined in Eqs. (\ref{eq:A}) and (\ref{eq:B}). It is evident that the sum rule $A+B=1$ is indeed satisfied.
}
\label{fig:ab}
\end{figure}

In Fig. \ref{fig:ab} we show the behavior of the two parameters $A$ and $B$ as a function of the interaction strength $g$. For convenience, in the following we rename the interaction strength as $g \to g'\equiv\hbar\omega_{0}\sigma_{0} g$, with the newly defined $g$ being dimensionless. This figure demonstrates that $A$ and $B$ behave as expected in the noninteracting ($g\lesssim0.01$) and Thomas-Fermi ($g\gtrsim100$) limits, and smoothly interpolate between the two. Also, it clearly shows that the sum rule $A+B=1$ is indeed satisfied. 

\begin{figure}[bp]
\centerline{\includegraphics[width=0.98\columnwidth]{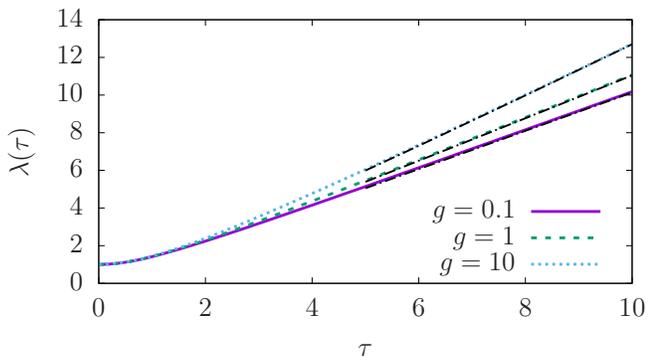}}
\caption{Evolution of the scaling parameter $\lambda(\tau)$ for different values of the interaction strength $g$.
The dot-dashed lines correspond to the asymptotic behavior obtained from Eq. (\ref{eq:asympt}).}
\label{fig:scaling}
\end{figure}

Let us now turn to the evolution of the system. We shall consider the solutions of Eq. (\ref{eq:scaling}), with the initial conditions $\lambda(0)=1$, $\dot\lambda(0)=0$, corresponding to a free expansion. In order to simplify the notations we introduce the dimensionless time coordinate $\tau\equiv\omega_{0}t$, and we also use the fact that  $A=B-1$. Then, we notice that the solution of the above scaling equation can be written in explicit form only in the noninteracting limit (where $B=0$), corresponding to the well-known  result $\lambda(\tau)=\sqrt{1+\tau^{2}}$ \cite{dalfovo1999}. When $B\neq0$, it is convenient to invert the relation between $\lambda$ and $\tau$, and transform Eq. (\ref{eq:scaling}) into an equation for $\tau(\lambda)$ \footnote{This can be done by introducing the auxiliary variable $u(\lambda)\equiv\dot\lambda(\tau)$, integrating with respect to $\lambda$, and taking into account that $\lambda(0)=1$, $\dot\lambda(0)=0$.}
\begin{equation}
\left(\frac{d\tau}{d\lambda}\right)^{2}
= \frac{\lambda^{2}}{(\lambda-1)\left[B(\lambda-1)+\lambda+1\right]} 
\label{eq:scaling-inv}
\end{equation}
the solution of which can be computed explicitly, but not inverted \footnote{Wolfram Research, Inc., Mathematica, Version 9, Champaign, IL (2013)}. Nevertheless, the asymptotic behavior for $\tau\gg1$ can be obtained first computing it for $\tau(\lambda)$, and then inverting. This yields
\begin{equation}
\lambda(\tau)\sim\frac{B}{B+1} W\left(\frac{1}{2 B} e^{\displaystyle \frac{(B+1)^{3/2}}{B}\tau+1}\right),
\label{eq:asympt}
\end{equation}
where $W(z)$ is the principal solution for the Lambert-W function \footnote{The Lambert-W function is defined as the inverse of $f(W)=We^{W}$. It is also called product logarithm (or omega function), see e.g. \href{http://mathworld.wolfram.com/LambertW-Function.html}{http://mathworld.wolfram.com/LambertW-Function.html}}.
The behavior of the scaling parameter $\lambda$ as a function of $\tau$, obtained from the numerical solution of Eq. (\ref{eq:scaling}), is shown in Fig. \ref{fig:scaling} for some values of $g$, along with their respective asymptotic limit (from Eq. (\ref{eq:asympt})).

\begin{figure}[tbp]
\centerline{\includegraphics[width=0.98\columnwidth]{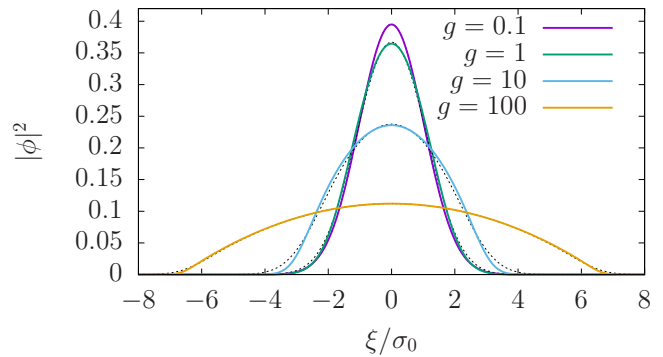}}
\caption{Comparison of the density $|\phi(\xi,\tau)|^{2}$ at $\tau=0$ (dashed lines) and $\tau=10$ (solid lines), for different values of the interactions, $g=0.1,1,10,100$. In all cases the fidelity $F(\tau)$ (see text) is greater that $0.99$. The density is plotted in units of $1/\sigma_{0}$.}
\label{fig:density}
\end{figure}

In the following we shall prove that the scaling approach discussed so far is indeed very effective in reproducing the exact dynamics of the system, as dictated by the time-dependent Gross-Pitaevskii equation (\ref{eq:gpe}).
 For practical purposes it is convenient to rewrite the latter in terms of the rescaled coordinates as 
\begin{equation}
i\hbar\partial_{t}\phi=\left[ -\frac{1}{\lambda^{2}}\frac{\hbar^2}{2m}\partial_{\xi}^{2} 
+\left(\frac{A}{\lambda^{2}}  + \frac{B}{\lambda}\right)\omega_{0}^{2}\xi^{2} + \frac{g'}{\lambda}|\phi|^{2} \right]\phi,
\end{equation}
with \cite{castin1996,kagan1996}
\begin{equation} 
\label{eq:scaling-psi}
\psi(x,t)=\frac{1}{\sqrt{\lambda(t)}}\phi\left(\xi,t\right)e^{\frac{im}{2\hbar}
\lambda(t)\dot\lambda(t)\xi^{2}},
\end{equation}
and $\lambda$ obtained from the numerical solution of Eq. (\ref{eq:scaling}). This greatly simplifies the calculations, as one needs to describe only the minor deviations of the rescaled wave functions with respect to the initial profile, the major effect of the expansion being captured by the scaling parameter. In the following we shall consider the evolution until a final time $\tau_{f}=10$, which is already well into the asymptotic regime (see Fig. \ref{fig:scaling}), and of the order of typical experimental times \cite{dalfovo1999}. In Fig. \ref{fig:density} we show the comparison of the density $|\phi(\xi,\tau)|^{2}$ at initial and final times, for different values of the interactions. 

\begin{figure}[tbp]
\centerline{\includegraphics[width=0.98\columnwidth]{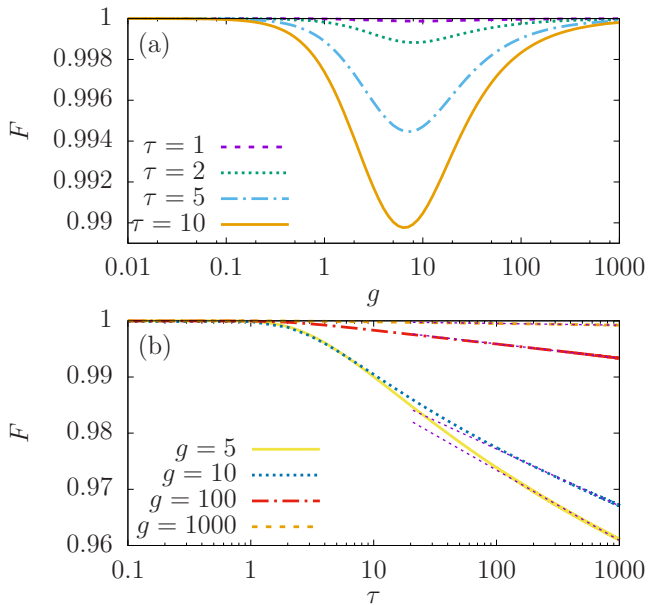}}
\caption{(a) Fidelity $F(\tau)=|\langle\phi(0)|\phi(\tau)\rangle|$ as a function of the interaction strength $g$, at different evolution times. (b) Fidelity $F(\tau)$ as a function of time, for different values of the interaction strength $g$. The thin dashed lines represent the asymptotic behavior that scales as $-\log(\tau)$.}
\label{fig:fidelity}
\end{figure}

Remarkably, the deviations from the exact behavior are negligible for almost any value of the interaction, with a fidelity $F(\tau)\equiv|\langle\phi(0)|\phi(\tau)\rangle|$ always greater than $0.99$ for $\tau\le\tau_{f}$, as shown in Fig. \ref{fig:fidelity}a. Obviously, the maximal deviations take place in the intermediate interaction regime. Fig. \ref{fig:fidelity}b shows that for larger times the fidelity decreases as $\log(\tau)$, and it remains close to unity within a few percent even for $\tau=10^{3}$ (a time that is well beyond the typical expansion times in the experiments).

\section{Conclusions}
We have considered an effective scaling approach for the free expansion of a one-dimensional interacting quantum wave packet, which is initially prepared in the ground state on a harmonic trap. The approach consists in looking for solutions that evolve self-similarly, at least \textit{on average}, so that the expansion of the system can be described solely by one scaling parameter which satisfies an ordinary differential equation. A direct comparison with the exact solutions of the Gross-Pitaevskii equation shows that the scaling approach is indeed very accurate for arbitrary values of the interactions. This result represents a proof-of-concept of the effectiveness of the scaling approach, which has been used by several authors in different situations \cite{guery-odelin2002,menotti2002,hu2003}, but never compared to the exact evolution. The present approach can be straightforwardly extended to higher dimensions.

\begin{acknowledgments}
M.M. acknowledge support by the Spanish Ministry of Economy, Industry and Competitiveness (MINECO) and the European Regional Development Fund FEDER through Grant No. FIS2015-67161-P (MINECO/FEDER, UE), and the Basque Government through Grant No. IT986-16. G.P. is supported by the Basque Government through the BERC 2014-2017 program and by MINECO through BCAM Severo Ochoa excellence accreditation SEV-2013-0323 and through Project No. MTM2016-76016-R ``MIP''. M.V. is supported by MINECO Grant No. FPA2015-64041-C2-1-P and by the Basque Government Grant No. IT979-16.
\end{acknowledgments}

%

\end{document}